\documentclass[a4paper,11pt]{article}
\pdfoutput=1 
\usepackage{aas_macros}
\usepackage{jcappub} 
\usepackage[T1]{fontenc} 
\usepackage{amsbsy}
\usepackage{color}
\usepackage{xcolor}
\usepackage{subcaption}
\usepackage{hyperref}
\usepackage{fontawesome}
\usepackage{multicol}

\usepackage[normalem]{ulem}

\newcommand{\be}{\begin{equation}}
\newcommand{\ee}{\end{equation}}

\usepackage{amssymb}

\title{\boldmath A cosmic degeneracy story: structure formation  with warm dark matter and scale-dependent primordial non-Gaussianities}

\author[a]{Cl\'ement Stahl,}
\author[a]{Benoit Famaey,}
\author[a]{Rodrigo Ibata,}
\author[a]{Katarina Kraljic,}
\author[a]{Fabien Castillo}

\affiliation[a]{Universit\'e de Strasbourg, CNRS, Observatoire astronomique de Strasbourg, UMR 7550, 67000 Strasbourg, France}

\emailAdd{clement.stahl@unistra.fr}

\abstract{It has been recently shown that cosmological models with scale-dependent primordial non-Gaussianities (sPNG) could provide a possible path to solve current cosmic tensions. Moreover, it has been pointed out that some of these models might mimic the effects of Warm Dark Matter (WDM) for several observables at low redshift. Here, we confirm the qualitative similarity of the matter power spectrum for sPNG and WDM models, but also point out differences in the halo mass function and void size function. We then jointly simulate WDM and sPNG together. Such simulations allow us to demonstrate that the joint impact of WDM and sPNG is close to the linear superposition of their respective effects at low redshift, at the percent level. We finally propose a model with mixed hot and cold dark matter together with sPNG, that reproduces the $\Lambda$CDM power spectrum at redshifts $z \leq 3$ but is still distinct in terms of halo statistics.}

\begin{document}

\maketitle

\flushbottom

\section{Introduction}\label{sec:Introduction}
The $\Lambda$CDM cosmological model is the current best parametric description of all measured cosmological observables. However, some existing discrepancies at the level of the inference of the parameters, if not due to systematics, may point toward new physics beyond $\Lambda$CDM \cite{Abdalla}. This cosmological model also provides the standard framework for the formation of galaxies, where several tensions are under scrutiny too \cite{Famaey, Bullock}. Possible solutions to such problems (both at cosmological and galactic scales) are not necessarily mutually exclusive \cite{Peebles:2020bph}, thus inducing possible degeneracies. 

One well-known alternative to $\Lambda$CDM, that is as old as the concept of Cold Dark Matter (CDM) itself, is to alter the mass of the fermionic particle DM candidates. This is known as Warm Dark Matter (WDM). Particle candidates for WDM are usually separated into thermal relics and non-resonantly produced sterile neutrinos in the keV to sub-keV range. The latter are in principle very natural candidates, for DM in general, and for WDM in particular. Adopting WDM allows for instance to naturally increase core radii in dwarf galaxies without resorting to feedback \cite{Bode:2000gq}, or to decrease the power spectrum at small scales, which might be desirable in view of, e.g., solving the $S_8$ cosmic tension \cite{Amon:2022azi, Brinckmann:2022ajr, Rubira:2022xhb,He:2023dbn,Chen:2024vuf}, or to make cosmic voids more empty \cite{Baldi:2024okt} than in $\Lambda$CDM. High redshift Lyman-alpha (Ly-$\alpha$) constraints at $z \gtrsim 4$ \cite{Viel:2013fqw} however strictly rule out WDM models with such observable consequences, but some aspects of their desirable properties can actually be kept -- while evading such high-redshift constraints --- when considering mixed DM models with a fraction of warm (or even hot) DM \cite{Boyarsky:2008xj,LoVerde:2014pxa,Parimbelli:2021mtp,Peters:2023asu,Kim:2023onk, Tan:2024cek,Pierobon:2024kpw,Dror:2024ibf}. Recently scale-dependent primordial non-Gaussianities (sPNG) \cite{Becker:2010hx,Becker:2012yr} were put forward as a possible very different alternative to ease cosmic tensions such as the $S_8$ one \cite{Stahl:2024stz}, with also possible consequences for small-scales problems \cite{Stahl:2022did, Stahl:2023ccv, Stahl:2024jzk}. As noted in Ref.~\cite{Baldi:2024okt}, such sPNG, whilst a very different alternative to $\Lambda$CDM than WDM or mixed DM models -- since it does not touch to the nature of DM itself -- can actually present a degeneracy with those approaches, at least at the level of the matter power spectrum. As in other works investigating cosmic degeneracies \cite{Baldi:2013iza,Baldi:2016oce,Hashim:2018dek} (see also \cite{Hashim:2014rda}), sPNG with WDM are also not mutually exclusive, and such a combination has never been attempted.

In this paper, we thus take a step further and simulate, for the first time, WDM and sPNG jointly. In Section \ref{sec:setups}, we describe the simulation codes and setups.  We then present in Section~\ref{sec:resWDM} our simulations of the effects of a thermal relic WDM particle with mass of $0.2$~kev (or sterile neutrino mass of 0.5~keV), along with sPNG. While -- as already pointed out hereabove -- such WDM models are actually ruled out by high-redshift Ly-$\alpha$ constraints \cite{Viel:2013fqw}, they are studied here as a benchmark to compare to sPNG. This will allow us to investigate whether the non-linear matter density field differs from the simple linear superposition of WDM and sPNG, and which degeneracies exist between these two $\Lambda$CDM modifications. We then investigate in Section~\ref{sec:resCWDM} a mixed DM model, with a small Hot Dark Matter (HDM) fraction, and sPNG. This is motivated by Ref.~\cite{Rogers:2023upm} who used such a model (without sPNG) to explain $z=3$ Ly-$\alpha$ observations. We also exhibit a case where the $\Lambda$CDM power spectrum at redshifts $z \leq 3$ is fully mimicked by a model with CDM, HDM and sPNG all simulated together. We finally conclude in Section \ref{sec:ccl}.

\section{Simulations codes and setups}\label{sec:setups}
The initial conditions of our numerical setup are generated using a modified version of the public code {\sc monofonIC} \href{https://bitbucket.org/ohahn/monofonic/}{\faGithub} \cite{Michaux:2020yis}, customized to take into account the scale-dependence of the PNG, expressed as follows \cite{Stahl:2024stz}: 
\begin{equation}
\label{eq:fNLk}
f_{\rm NL}(k)=\frac{f_{\rm NL}^0}{1+\alpha}\left[\alpha+\tanh \left(\frac{k-k_{\rm min}}{\sigma} \right) \right],
 \end{equation}
In order to have a better suppression of the PNG at large scales, we take in this work $\alpha=\tanh\left(\frac{k_{\rm min}}{\sigma} \right)$ instead of $\alpha=1$, which does not affect the main results presented in Ref.~\cite{Stahl:2024stz}. To generate the initial conditions, {\sc Monofonic} interfaces with the Boltzmann solver {\sc CLASS} \href{https://lesgourg.github.io/class_public/class.html}{\faGithub} \cite{Lesgourgues:2011rh}. It takes as an input \texttt{m}\_{\texttt{ncdm}} that is the sterile neutrino mass. It can be converted to an equivalent thermal relic mass using eg.~Eq.~(6) of Ref.~\cite{Peters:2023asu}. A sterile neutrino mass of 0.5~keV corresponds to a thermal relic mass of 0.2~keV. To numerically follow the DM density field down to $z=0$, we use the public code {\sc Gadget-4} \href{https://wwwmpa.mpa-garching.mpg.de/gadget4/}{\faGithub} \cite{Springel:2020plp}. Table \ref{tab:cosmopar} presents the main cosmological parameters used for all our simulations. Table \ref{tab:numparam} describes the parameters common to every run, while in Table \ref{tab:eachSimu} we exhibit the parameters specific to each run.

\begin{table*}
\begin{center}
\caption{Main cosmological parameters used in this work: matter density $\Omega_{\rm m}$, baryonic density $\Omega_{\rm b}$, dark energy density $\Omega_{\Lambda}$, Hubble constant $H_0$, amplitude of the primordial power spectrum $A_S$, and spectral index $n_s$.}
\begin{tabular}{ |c| c | c | c | c |  c | c |}
\hline
 $\Omega_{\rm m}$ & $\Omega_{\rm b}$ & $\Omega_{\Lambda}$ & $H_0$  & $A_s$ & $n_s$    \\
 & & & [km/s/Mpc] & & \\
\hline 

 0.31 & 0.049 & 0.69 & 67.7 & $2.11 \times 10^{-9}$ & 0.967 \\

\hline
\end{tabular}
\label{tab:cosmopar}
\vspace{-5mm}
\end{center}
\end{table*}

\begin{table*}
\begin{center}

\caption{Main numerical parameters used in this work: box length $L_{\rm box}$, number of particles $N_{\rm part}$ , starting redshift $z_{\rm start}$, particle masses $m_{\rm part}$, and softening length $L_{\rm soft}$.}
\begin{tabular}{ |c| c | c | c |  c |}
\hline  $L_{\rm box}$  & $N_{\rm part}$  & $z_{\rm start}$ & $m_{\rm part}$ &  $L_{\rm soft}$ \\ 

 [Mpc/$h$] &  & &  [$M_{\odot}$/$h$] &  [kpc/$h$]\\
\hline 

 500 & $512^3$ & 32 & $8 \times 10^{10}$ &  50 \\

\hline
\end{tabular}
\label{tab:numparam}
\vspace{-5mm}
\end{center}
\end{table*}

\begin{table*}
\begin{center}

\caption{Specifics of each simulation: CDM density $\Omega_{\rm cdm}$, WDM density $\Omega_{\rm wdm}$, sterile neutrino mass $m_{\rm wdm}$, WDM (or HDM) fraction $f_{\rm wdm}$, amplitude of the sPNG, $f_{\rm NL}^0$, and parameters governing its shape (from Eq.~\ref{eq:fNLk}) $\left\{\sigma,\,k_{\rm min}\right\}$. All other cosmological parameters are kept as in Table~\ref{tab:cosmopar}.}
\begin{tabular}{|l ||c| c | c |c | c |  c | c|}
\hline 
  & $\Omega_{\rm cdm}$  & $\Omega_{\rm wdm}$  & $m_{\rm wdm}$ & $f_{\rm wdm}$ & $f_{\rm NL}^0$ & 
 $\sigma$ & $k_{\rm min}$ \\

Simulation  &  & &  [ev] & \% & & [$h$/Mpc] & [$h$/Mpc]\\
\hline 

 CDM & 0.26 & 0 & 0 & 0 &  0 & 0 & 0 \\
  \hline
  WDM & 0 & 0.26 & 500 & 100 &  0 & 0 & 0 \\
    mixed DM & 0.255 & 0.005 & 10 & 2 & 0 & 0 & 0 \\
    \hline
  $f_{\rm NL}^0=-500$ & 0.26 & 0 & 0 & 0 & -500 & 0.1 & 0.15 \\
  $f_{\rm NL}^0=500$ & 0.26 & 0 & 0 & 0 & 500 & 0.1 & 0.15 \\
  \hline
  $f_{\rm NL}^0=-500$\,\&\,WDM & 0 & 0.26 & 500 & 100 &  -500 & 0.1 & 0.15 \\
    $f_{\rm NL}^0=500$\,\&\,WDM & 0 & 0.26 & 500 & 100 & 500 & 0.1 & 0.15 \\
    \hline
      $f_{\rm NL}^0=-500$\,\&\,mixed DM & 0.255 & 0.005 & 20 & 2 &  -500 & 0.1 & 0.15 \\
    $f_{\rm NL}^0=500$\,\&\,mixed DM & 0.255 & 0.005 & 20 & 2 & 500 & 0.1 & 0.15 \\
\hline
\end{tabular}
\label{tab:eachSimu}
\vspace{-5mm}
\end{center}
\end{table*}
\section{Scale-dependent PNG and WDM}\label{sec:resWDM}

In this section, we compare the benchmark Gaussian CDM case to five alternative simulations: first, a WDM simulation with sterile neutrino mass of $500$~eV, then two simulations with sPNG (with positive or negative $f_{\rm NL}$ at small scales), and finally two simulations with joint WDM and sPNG (with positive or negative  $f_{\rm NL}$ at small scales). In all cases, the total dark matter density and dark energy density are both untouched compared to the benchmark CDM model. We present three different diagnostics to compare those simulations: the matter power spectrum, the Halo Mass Function (HMF) and the Void Size Function (VSF).

\subsection{Matter power spectrum}

               \begin{figure}

        \includegraphics[width=\textwidth]{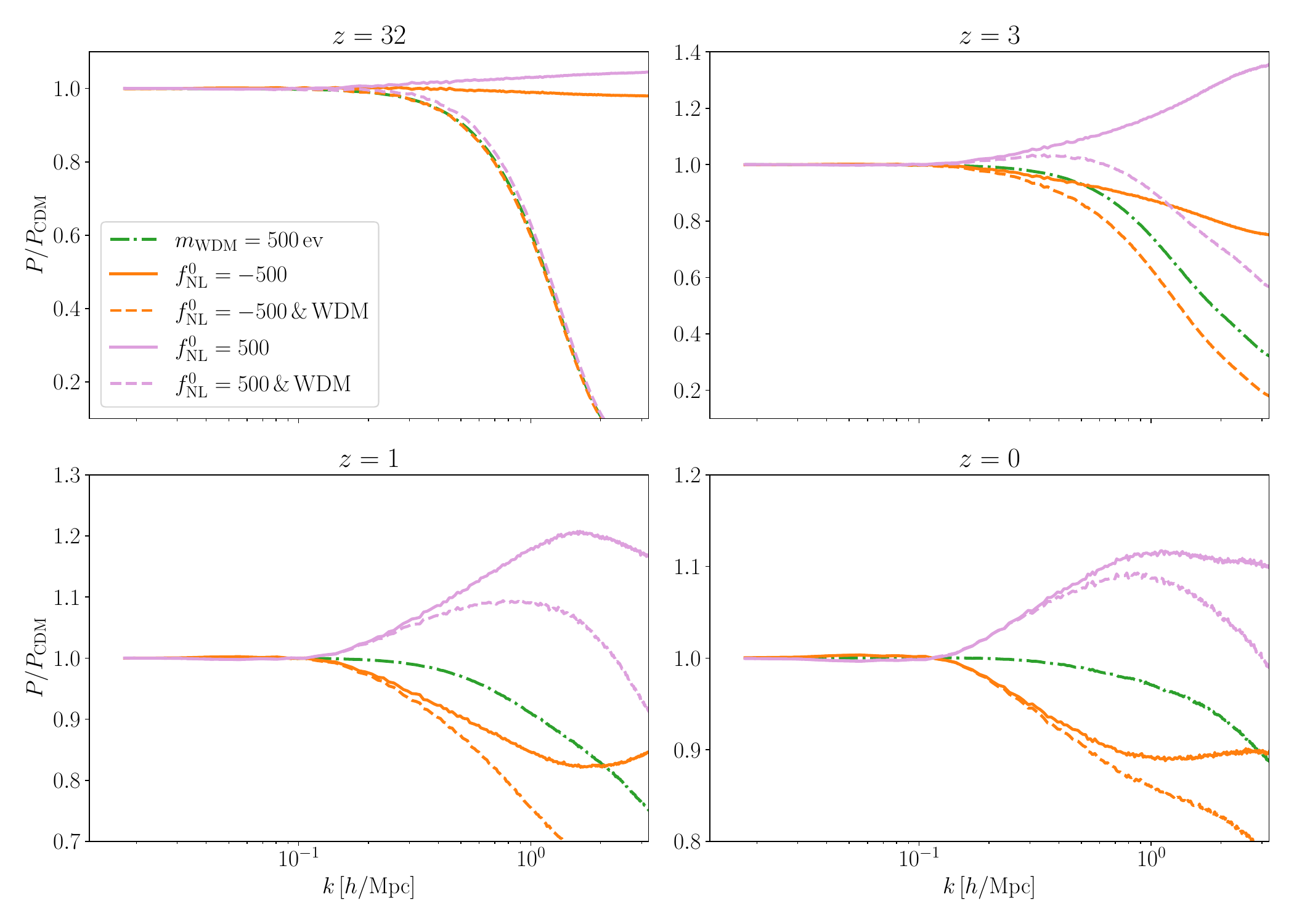}
                \caption{Ratio of the matter power spectrum to the benchmark Gaussian CDM case, for the five different simulations studied in Sect.~\ref{sec:resWDM} (cf.~Table \ref{tab:eachSimu}), at four redshifts ($z=32, 3, 1, 0$).}
                      \label{fig:PS_WDM}
         \end{figure}

In Fig.~\ref{fig:PS_WDM}, we present the ratio of the power spectrum of the five simulations to that of the CDM case at four different redshifts ($z=32,3,1,0$). At $z=32$, one can immediately see at high $k$ a sharp drop in the pure WDM power spectrum. It is related to the free-streaming scale, and this sharp drop explains why such models are actually ruled out by high-$z$ observations \cite{Viel:2013fqw}. A negative small-scale $f_{\rm NL}$, conversely, only affects very mildly the power spectrum at $z=32$, but progressively reaches an effect very similar to WDM at lower redshifts, showing explicitly why such models present a more promising tentative solution to the $S_8$ tension than pure WDM models. On the other hand, a positive small-scale $f_{\rm NL}$ tends to slightly boost the power spectrum at small scales, mildly so at $z=32$, and with a significant signature at $z=3$, which then decreases while reaching $z=0$. 

When considering joint WDM \& sPNG simulations, the two effects appear to almost linearly superpose (see Eq.~\eqref{eq:Psum}). Hence, all such models are ruled out at high $z$, where the WDM effect largely dominates. However, the two effects become comparable in amplitude at low $z$. In the negative $f_{\rm NL}$ case, the power spectrum drop has therefore a larger amplitude than in the two separate cases, whilst in the positive $f_{\rm NL}$ case, the power spectrum boost is tapered by WDM. This means that, it should in principle be possible to construct a model with WDM and positive small-scale $f_{\rm NL}$ that would produce the same low-$z$ power spectrum as $\Lambda$CDM. We will come back to this question in Sect.~\ref{sec:resCWDM}.

                        \begin{figure}

        \includegraphics[width=\textwidth]{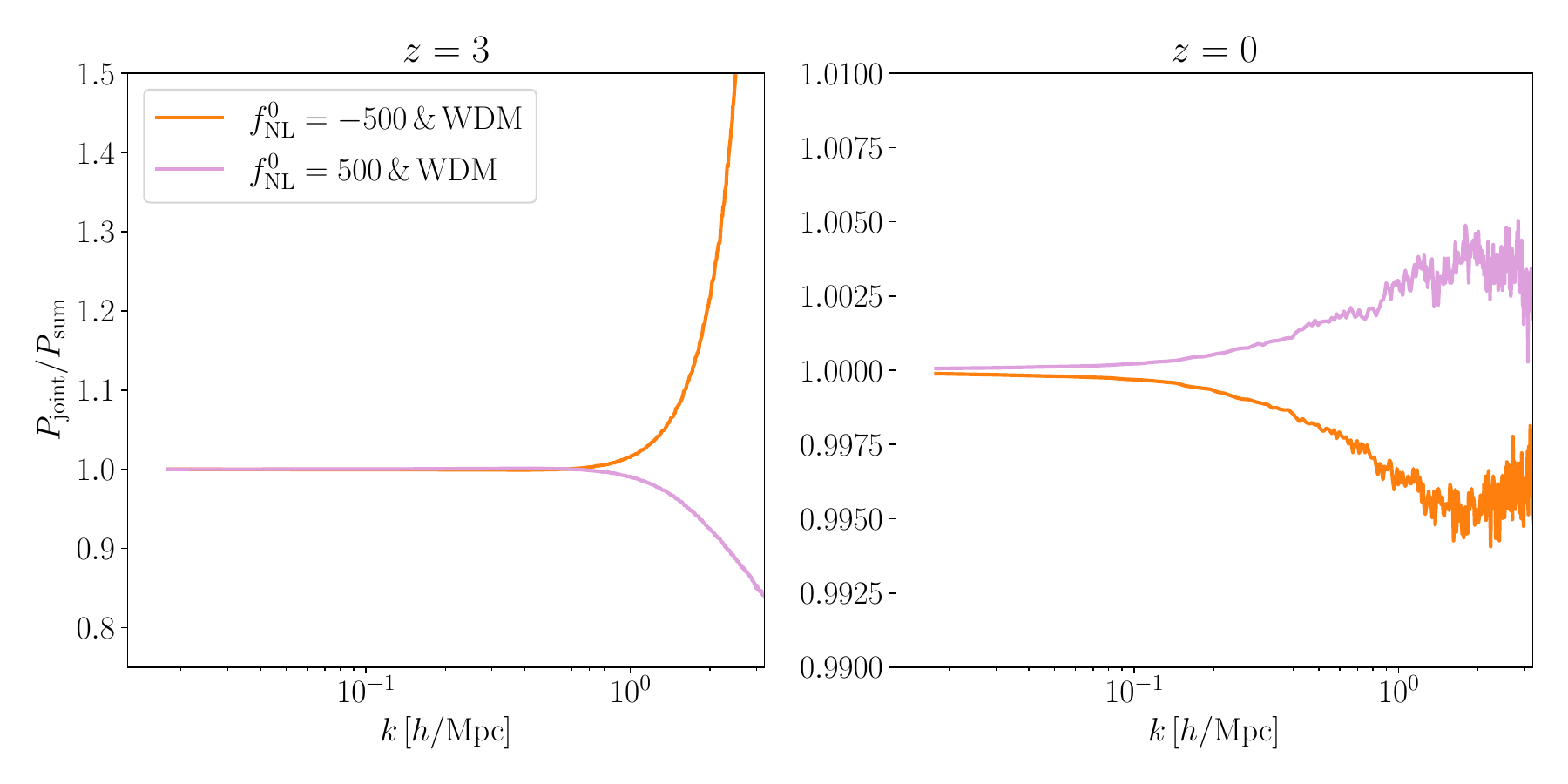}
                \caption{Ratio of the joint (WDM and sPNG simulated together) matter power spectrum $P_{\rm joint}$ with respect to the linear combination $P_{\rm sum}$ described in Eq.~\eqref{eq:Psum}. At $z=3$, $P_{\rm sum}$ is an excellent approximation down to $k=1$, whilst at $z=0$ it approximates $P_{\rm joint}$ at less than percent-level accuracy.}
                      \label{fig:PS_jointvssum}
         \end{figure}

Let us now quantify whether the joint effect of WDM \& sPNG on the power spectrum is indeed well approximated by a linear superposition of each effect. Note that, since we are considering non-linear scales, this is far from a trivial statement. Let us consider the modified power spectra (compared to the CDM case) as $P_{\rm WDM} = P_{\rm CDM} + \varepsilon_{_{\rm WDM}}$ and $P_{\rm PNG} = P_{\rm CDM} + \varepsilon_{_{\rm PNG}}$, and the linear superposition as 
         \begin{equation}
         \label{eq:Psum}
             P_{\rm sum} \equiv P_{\rm CDM}+ \varepsilon_{_{\rm WDM}} + \varepsilon_{_{\rm PNG}}=P_{\rm WDM}+P_{\rm PNG}-P_{\rm CDM}.
         \end{equation}
In Fig.~\ref{fig:PS_jointvssum}, we plot the ratio of the power spectrum of the joint WDM \& sPNG over the linearly superposed power spectrum $P_{\rm sum}$. At $z=3$, the two are very similar down to $k=1$, but the linear superposition overestimates the true effect of sPNG at smaller scales, by up to 50\%. The difference is much smaller at larger $z$, but more surprisingly the difference is also much smaller at lower $z$. At $z=1$ and $z=0$, the $P_{\rm sum}$ approximation is accurate at less than 1\%.

\subsection{Halo Mass function}

               \begin{figure}

        \includegraphics[width=\textwidth]{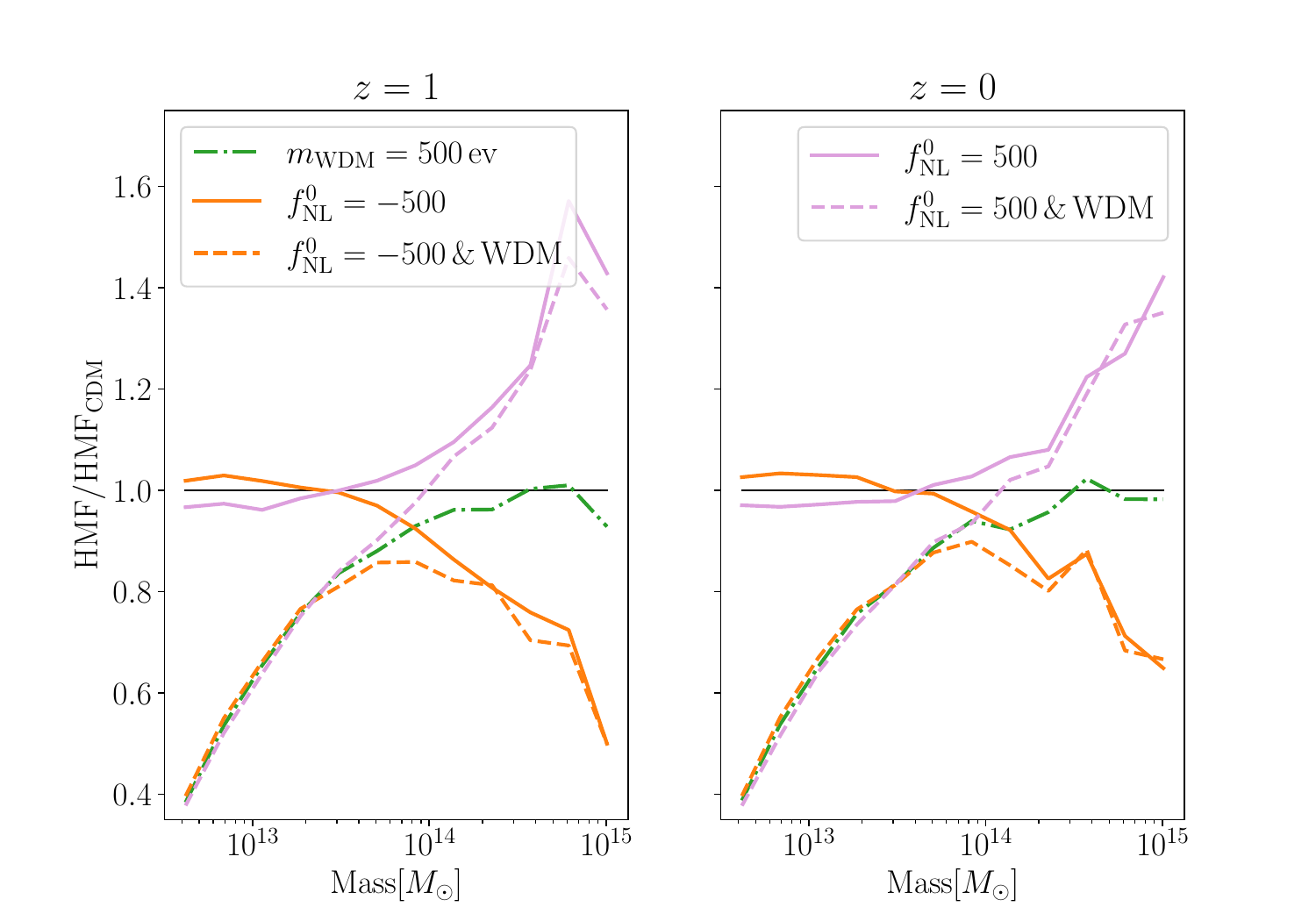}
                \caption{Ratio of the Halo Mass Function (HMF) to the one in the benchmark Gaussian CDM case for the five different simulations studied in Sect.~\ref{sec:resWDM}, at redshifts $z=1$ and $z=0$.}
                      \label{fig:HMF_WDM}
         \end{figure}

Let us now consider the effect of WDM and sPNG, first simulated separately then together, on the HMF at different redshifts. We identified halos using {\sc subfind} \cite{Springel:2000qu}, which is included in the public version of {\sc Gadget-4} and builds the halos catalog on the fly while the simulation runs. We have shown hereabove that, despite having different signatures at high $z$, WDM or a negative small-scale $f_{\rm NL}$ impact the power spectrum at low $z$ in a very similar fashion. This is {\it not} the case for the HMF. Indeed, as illustrated on Fig.~\ref{fig:HMF_WDM}, WDM depletes the low-mass end of the HMF whilst a negative $f_{\rm NL}$ depletes the high-mass end. A positive $f_{\rm NL}$, on the other hand, boosts the high-mass end of the HMF with respect to the CDM case: a different $\sigma_8$ could make it match the CDM case at high mass and lead to less small halos, as in WDM. But in that case, the HMF would be constant at low mass without any WDM-like low-mass cut-off, as illustrated by the flattening of the positive $f_{\rm NL}$ curve at low mass compared to the steeply decreasing WDM one on Fig.~\ref{fig:HMF_WDM}.

In turn, joint WDM \& sPNG simulations (dashed lines on Fig.~\ref{fig:HMF_WDM}) either deplete both ends of the HMF in the case of negative $f_{\rm NL}$, hence increasing the concavity of the HMF, or flatten the HMF in the case of positive $f_{\rm NL}$. Therefore, for the most virialized objects of the simulation box, no degeneracy with the CDM case is expected at low redshift despite possibly having the same statistical properties at the level of the matter power spectrum. Interestingly, summing the separated HMFs in the spirit of Eq.~\eqref{eq:Psum} allows to recover the HMF of the joint WDM \& sPNG simulations at the 5\% level.

\subsection{Void Size Function}

               \begin{figure}

        \includegraphics[width=\textwidth]{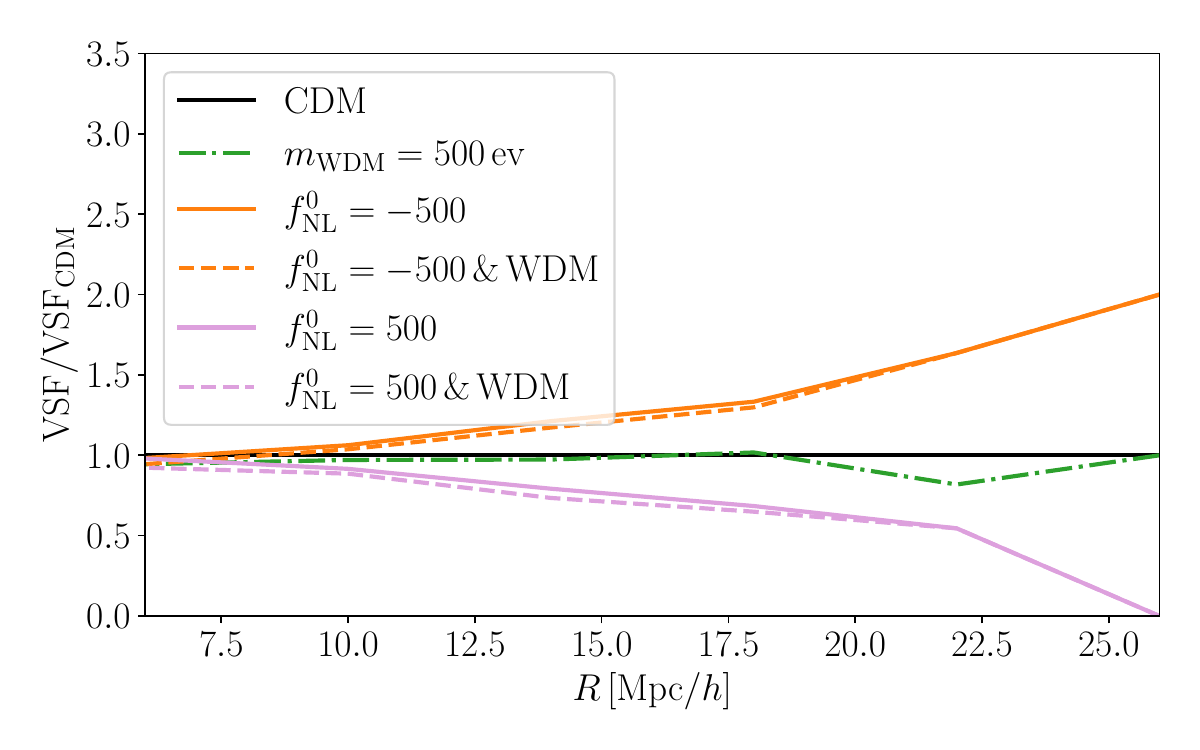}
                \caption{At redshift $z=0$, ratio of the Void Size Function (VSF) to the benchmark Gaussian CDM case, for the five different simulations studied in Sect.~\ref{sec:resWDM}.}
                      \label{fig:VSF_WDM}
         \end{figure}
         
To further explore the similarities and differences between sPNG and WDM, let us now consider the least virialized `objects' of the simulations, namely voids and the Void Size Function (VSF). We identified the voids using the void finder of \texttt{Pylians} with an underdensity threshold of $\delta_{\rm t}=-0.7$. On Fig.~\ref{fig:VSF_WDM}, we plot the ratio of the VSF in our 5 simulations with respect to that in the benchmark Gaussian CDM case. As is well known (eg.~\cite{Yang:2014upa}), WDM does not affect much the VSF hence leads, in our numerical setup, to a constant ratio of $\sim 1$. With negative $f_{\rm NL}$, the VSF rises for large voids, whilst it decreases for large voids for positive $f_{\rm NL}$. In the joint WDM \& sPNG simulations, the effect is exactly the same as in the sPNG simulations with CDM. Therefore, for the least virialized objects of the simulation box, no degeneracy with the CDM case is expected for the WDM \& sPNG case at low redshift, despite possibly having the same statistical properties at the level of the matter power spectrum.

\section{Scale-dependent PNG and mixed HDM/CDM}\label{sec:resCWDM}

The matter power spectrum of Fig.~\ref{fig:PS_WDM} presents at $z=32$ a sharp drop at high $k$ that observationally rules out all WDM models explored hereabove. To keep some aspects of the desirable properties of WDM while evading high-$z$ constraints, we now consider a mixed DM model with a fraction of hot DM, which we couple to sPNG. For instance, motivated by the suppressed small-scale matter power spectrum amplitude and tilt measured from the eBOSS Ly-$\alpha$ forest at $z=3$, a model with a 2\% fraction of hot DM with 10~eV mass has been put forward in Ref.~\cite{Rogers:2023upm}. We base our following investigations on coupling this model with sPNG. 

               \begin{figure}

        \includegraphics[width=\textwidth]{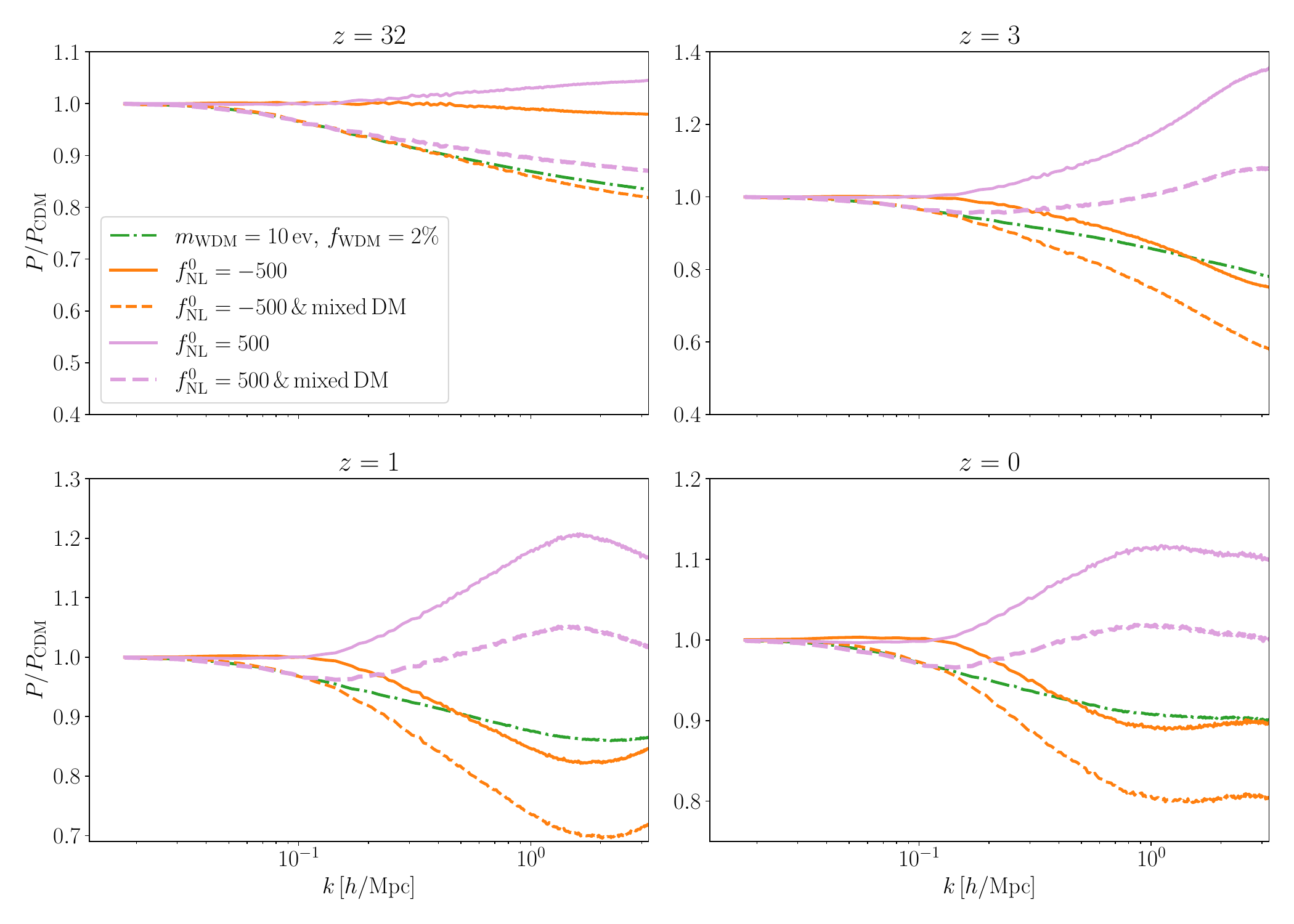}
                \caption{Ratio of the matter power spectrum to the CDM case, for the three mixed DM simulations studied in Sect.~\ref{sec:resCWDM} along with the two sPNG simulations already studied in Section \ref{sec:resWDM} (cf.~Table \ref{tab:eachSimu}), at four redshifts ($z=32, 3, 1, 0$).}
                      \label{fig:PS_HDM}
         \end{figure}

                                 \begin{figure}

        \includegraphics[width=\textwidth]{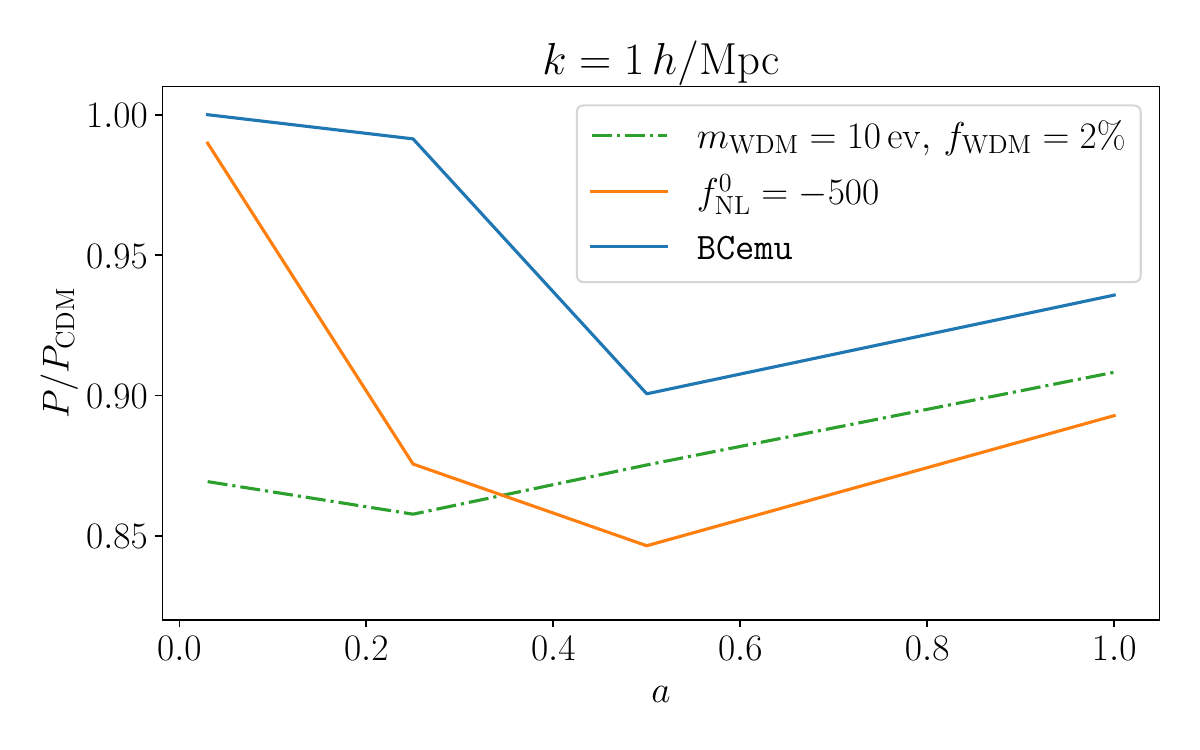}
                \caption{Ratio of the matter power spectrum to the benchmark Gaussian CDM case as a function of scale factor $a$ for wavenumber $k=1\,h/$Mpc. The three different models considered are the sPNG model with negative $f^0_{\rm NL}=-500$ and CDM, the (Gaussian) mixed DM model, and a model with baryonification using  the emulator \texttt{BCemu}. The models have a different temporal development thus allowing to distinguish their effects on cosmological observables.}
                      \label{fig:zDEP}
         \end{figure}

In Fig.~\ref{fig:PS_HDM}, we present the ratio of the power spectrum of our three new simulations (where we swap WDM for mixed DM) to that of the benchmark Gaussian CDM case at the same four different redshifts ($z=32,3,1,0$) as above. We also show the sPNG models with CDM, for reference. At $z=32$, one can immediately see that the mixed HDM/CDM model does not suffer from the same free-streaming cutoff as the WDM case. The power spectrum is damped in a much less drastic way. Interestingly at $z \leq 3$, the damping of the power spectrum of the mixed DM model is again very similar to that of a negative small-scale $f_{\rm NL}$. This is illustrated on Fig.~\ref{fig:zDEP}, where the matter power spectrum at wavenumber $k=1 h$/Mpc is displayed as a function of the scale-factor $a$. The effect of a negative small-scale $f_{\rm NL}$, of a 2\% fraction of HDM, or from baryonic feedback\footnote{As in Ref.~\cite{Stahl:2024stz}, we consider the 7 vanilla values of the parameters governing the baryonic physics: five parameters for the gas: $M_c= 10^{13.3}$, $\mu=0.93$, $\theta_{\rm ej}=4.2$, $\gamma=2.25$, $\delta=6.4$, and two parameters about the stars: $\eta=0.15$ and $\eta_{\delta}=0.14$, see Table 1 of Ref.~\cite{Schneider:2018pfw} for more details.} modeled with the emulator \texttt{BCemu} \href{https://github.com/sambit-giri/BCemu}{\faGithub} \cite{Schneider:2018pfw,Giri:2021qin} are all very similar at redshifts $z \leq 1$ ($a \geq 0.5$). However, the dependence of the scale factor with redshift becomes very different at higher redshifts, since the baryon feedback case quickly goes back to the no-feedback case, while this happens at higher redshifts for the sPNG case, and never happens in the mixed DM case (although the damping of the power spectrum still passes CMB constraints in all cases).

When considering joint mixed DM \& sPNG simulations, the two effects again almost linearly superpose (following Eq.~\ref{eq:Psum}), at the 1-2\% level. In the negative $f_{\rm NL}$ case, the power spectrum displays therefore a larger drop than in the two separate cases, whilst in the positive $f_{\rm NL}$ case, the power spectrum boost is tapered by mixed DM.

               \begin{figure}

        \includegraphics[width=\textwidth]{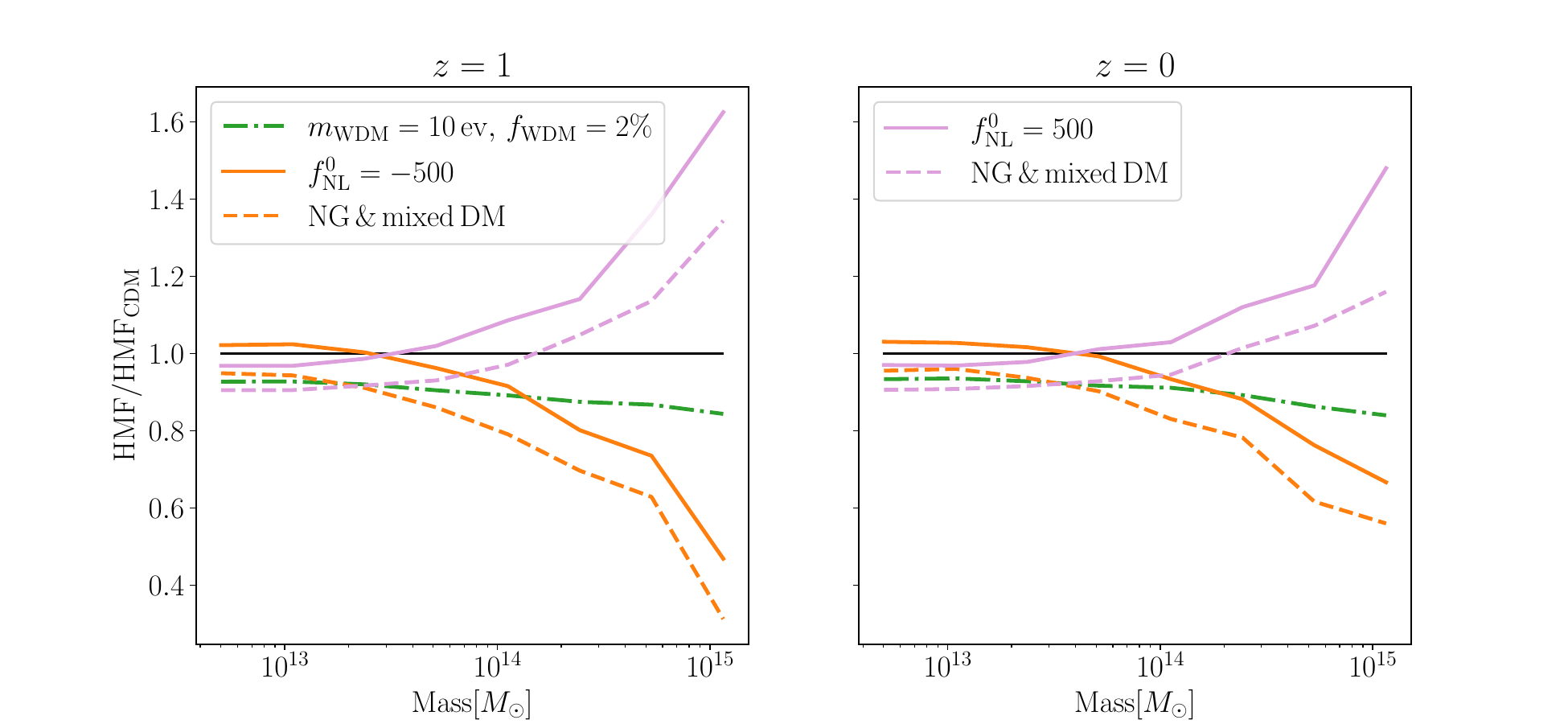}
                \caption{Ratio of the Halo Mass Function (HMF) to the one in the CDM case, for the three mixed DM simulations studied in Sect.~\ref{sec:resCWDM} along with the two sPNG simulations already studied in Section \ref{sec:resWDM}, at redshifts $z=1$ and $z=0$.}
                      \label{fig:HMF_HDM}
         \end{figure}

The latter model with mixed HDM/CDM and positive small-scale $f_{\rm NL}$ nicely illustrates the degeneracies between the different extensions of $\Lambda$CDM considered in the present paper, as this model, which is not ruled out at high redshift, produces a power spectrum agreeing very well with the CDM one at the 5\% level at  $z \leq 3$. However, as shown on Fig.~\ref{fig:HMF_HDM}, the degeneracy is broken at the level of the HMF at high masses due to the effect of sPNG, in line with our analysis in the previous section. 

This example effectively conveys the idea of Ref.~\cite{Peebles:2020bph} (see beginning of its Section 7) that a combination of sPNG and some hotter form of DM could pass existing cosmological constraints while subtly modifying the context of galaxy formation. Recognizing that WDM models face significant challenges from high-$z$ observations of the Ly-$\alpha$ forest, it was proposed that such problems could be addressed by introducing a mixed dark matter model as originally suggested by Ref.~\cite{Davis:1981yf} and still under discussion~\cite{Drewes:2016upu,Rogers:2023upm}. It was then suggested that such models could be coupled with PNG. Noting that the near-Gaussian nature of CMB anisotropies might seem to contradict the presence of such PNG, it was already pointed out that PNG could vary with scale. The simulations presented hereabove are prime examples of such models. The case with positive small-scale $f_{\rm NL}$ and mixed DM produce a power spectrum almost indistinguishable from the CDM case, but favors a fast overproduction of high mass halos which could be interesting in view of JWST detections of massive galaxies at high $z$ \cite[e.g.,][]{Biagetti:2022ode}. In view of exploring the consequences on galaxy formation, one would need to run zoom-in simulations in the spirit of Refs.~\cite{Nadler:2024ejs, An:2024mgq, Nadler:2024fcs}, but adding sPNG to the picture. On the other hand, the case with negative small-scale $f_{\rm NL}$ and mixed DM can help damping the non-linear power spectrum even more efficiently than either of those cases taken in isolation, thereby providing a possibly easy solution to the $S_8$ tension \cite{Peters:2023asu,Stahl:2024stz} without resorting to aggressive feedback.

\section{Conclusions and perspectives}\label{sec:ccl}

In this article, we have explicitly shown how sPNG models can replicate the effects of WDM or mixed HDM/CDM at the level of the power spectrum at low redshift. We have however shown that the $z$-dependency allows, in general, to distinguish both types of models. Interestingly, while the effects explored in this paper happen in the non-linear regime of structure formation, we have demonstrated that the joint effect of WDM \& sPNG (or mixed DM \& sPNG) is very well approximated by a simple linear superposition of their separate effects, at the level of the power spectrum as well as at the level of the HMF, accurate to percents level. As an illustration of the possible degeneracies, we have also presented a model with both mixed DM and sPNG, which produces a power spectrum almost {\it indistinguishable} from the CDM case (but favors an overproduction of high mass halos, which breaks the degeneracy). On the other hand, a negative small-scale $f_{\rm NL}$ jointly simulated with sPNG can help damping the non-linear power spectrum even more efficiently than either of those cases taken in isolation, thereby providing a possibly easy solution to the $S_8$ tension \cite{Peters:2023asu,Stahl:2024stz} without resorting to aggressive feedback. Such a solution to the $S_8$ tension may then possibly be combined to a linear solution to the $H_0$ tension, following e.g.~\cite{Vagnozzi:2023nrq,Toda:2024ncp,Simon:2024jmu}. The general lesson from the simulations presented in this paper is that there is still room for multiple extensions of $\Lambda$CDM simulated together that can mimic very precisely some $\Lambda$CDM observables while subtly modifying others, which is a helpful tip to keep in mind in view of current cosmic tensions. 

\acknowledgments
CS and BF acknowledge valuable input from Frédéric Bournaud. This work has made use of the Infinity Cluster hosted by the Institut d'Astrophysique de Paris.

\section*{Softwares}
The analysis was partially made using \texttt{YT} \href{https://yt-project.org/}{\faGithub} \cite{Turk:2010ah}, \texttt{Pylians} \href{https://pylians3.readthedocs.io/en/master/index.html}{\faGithub} \cite{Pylians} as well as IPython \cite{Perez:2007emg}, Matplotlib \cite{Hunter:2007ouj} and NumPy \cite{vanderWalt:2011bqk}.

\section*{Authors' Contribution}
This work is based on a master project started in February 2024 by Théo Bruant and FC under the supervision of CS. Subsequently CS performed the more systematic study described in this work in consultation with BF. CS and BF drafted the manuscript. All the authors improved it by their comments.

\section*{Carbon Footprint}
Following the calculations of Ref.~\cite{berthoud}, that account for the global utilization of a cluster and the pollution due to the electrical source: 1 hour core is equivalent to 4.7 gCO2e. 
The simulations presented in this work required then 0.5 TCO2eq.

\bibliographystyle{JHEP.bst}
\bibliography{ref.bib}

\end{document}